\documentclass[AMA,STIX1COL]{WileyNJD-v2}
\usepackage{bm}
\usepackage{setspace}
\articletype{Article Type}%

\received{26 April 20YY}
\revised{6 June 20YY}
\accepted{6 June 20YY}

\begin{document}

%\title{A Simulation-free Group Sequential Design with Max-combo Tests in the Presence of Non-proportional Hazards\protect\thanks{A Group-sequential Design for Max-combo tests}}
\title{A Simulation-free Group Sequential Design with Max-combo Tests in the Presence of Non-proportional Hazards}

\author[1]{Lili Wang}

\author[2]{Xiaodong Luo}

\author[2]{Cheng Zheng}

%\authormark{AUTHOR ONE \textsc{et al}}
\authormark{Wang, L.,Luo, X. and Zheng, C.}

\address[1]{\orgdiv{Department of Biotatistics}, \orgname{ University of Michigan, Ann Arbor}, \orgaddress{\state{Michigan}, \country{U.S.A.}}}

\address[2]{\orgdiv{Department of Biostatistics and Programming, Research and Development}, \orgname{Sanofi US, Bridgewater}, \orgaddress{\state{New Jersey}, \country{ U.S.A.}}}

%\address[3]{\orgdiv{Org Division}, \orgname{Org Name}, \orgaddress{\state{State name}, \country{Country name}}}

\corres{*Cheng Zheng, Department of Biostatistics and Programming, Research and Development, Sanofi US, Bridgewater, New Jersey, U.S.A.  \email{Cheng.Zheng@sanofi.com}}

%\presentaddress{This is sample for present address text this is sample for present address text}

\abstract[Summary]{Non-proportional hazards (NPH) have been observed in many immuno-oncology clinical trials. Weighted log-rank tests (WLRT) with suitable weights can be used to improve the power of detecting the difference between survival curves in the presence of NPH. However, it is not easy to choose a proper WLRT in practice. A versatile maxcombo test was proposed to achieve the balance of robustness and efficiency, and has received increasing attention recently. Survival trials often warrant interim analyses due to their high cost and long durations. The integration and implementation of maxcombo tests in interim analyses often require extensive simulation studies. In this report, we propose a simulation-free approach for group sequential designs with the maxcombo test in survival trials. The simulation results support that the proposed method can successfully control the type I error rate and offer excellent accuracy and flexibility in estimating sample sizes, with light computation burden. Notably, our method displays strong robustness towards various model misspecifications and has been implemented in an R package for free access online.}

\keywords{delayed treatment effect, group sequential design, interim analysis, maxcombo, non-proportional hazard, weighted log-rank test}

%\jnlcitation{\cname{%
%\author{Williams K.}, 
%\author{B. Hoskins}, 
%\author{R. Lee}, 
%\author{G. Masato}, and 
%\author{T. Woollings}} (\cyear{2016}), 
%\ctitle{A regime analysis of Atlantic winter jet variability applied to evaluate HadGEM3-GC2}, \cjournal{Q.J.R. Meteorol. Soc.}, \cvol{2017;00:1--6}.}

\maketitle

%\footnotetext{\textbf{Abbreviations:} ANA, anti-nuclear antibodies; APC, antigen-presenting cells; IRF, interferon regulatory factor}

\section{Introduction}
\label{s:intro}
Survival outcomes are common end-points of interest in confirmatory trials to demonstrate treatment effect in oncology. In the presence of non-proportional hazards (NPH), which has been increasingly encountered in practice, the detection power of the commonly used log-rank test is much lower than those under proportional hazards (PH). For instance, delayed treatment effect was often reported in immune-directed anti-cancer therapies 
\cite{reck2016pembrolizumab,mok2019pembrolizumab}. Unlike chemotherapy, which displays early antitumor effects or separation between the survival curves, immunotherapy stimulates the patient's immune system for an antitumor response, causing delayed clinical effects \cite{mick2015statistical}. 

Weighted log-rank tests (WLRT) incorporate time-dependent weights to improve the detection power in the presence of non-proportional hazards. Weight functions can be dependent on survival functions \cite{harrington1982class} or at-risk proportions \cite{gehan1965generalized,tarone1977distribution}. In this report, we focus on the Fleming-Harrington class of WLRT, but one can easily extend the proposed design to other weight functions. The shape of the Fleming-Harrington weight function can be adjusted according to the survival curves. For example, one can put more weight on late separation for delayed treatment effect to improve the detection power. According to Schoenfeld \cite{schoenfeld1983sample}, the optimal weight (with the highest power) should be proportional to the logarithm of the hazard ratio \cite{schoenfeld1981asymptotic}, which is a time-dependent function under NPH.

Survival curves are generally unknown, and thus, the appropriate weight function cannot be decided before starting the trial. Lee \cite{lee1996some} proposed a versatile max-combo test, which takes the maximum value of a set of different WLRT to provide a robust detection. In other words, whether it is PH or NPH, the max-combo test gives power quite close to the optimal one in the combo of different WLRT. A typical maxcombo test combines several WLRT, each of which is most powerful in detecting a certain pattern of NPH or PH difference between treatment arms, and the multiple testing adjustment is conducted via a Dunnett-type parametric method. %Moreover, the selected best weight reflects the exact hazard ratio between the two treatment arms.  

Interim analysis (IA) in group sequential design (GS) enables multiple looks (interims) before the end of the study. They will allow early stops when there is sufficient evidence to discontinue the study, like the rejection of the null hypothesis, toxic effects, and futility. Though their benefits have been extensively studied, to the best of our knowledge, group sequential designs with max-combo tests (GS-MC) have not been systematically established. In this report, we develop a simulation-free approach to calculate the stopping boundaries with GS-MC design. As shown in the sequel, our methods can control type I error and provide an efficient way of computing the power and sample size with a GS-MC design. 

The rest of this report is organized as follows.  We introduce the designs of WLRT and the max-combo test, and extend them to GS-MC in Section~\ref{s:problem}. We propose simulation-free approaches for GS-MC to compute boundaries for type I error control and practical sample size in realistic scenarios in Section~\ref{s:solution}. We evaluate the proposed methods through extensive simulations with or without violations of the model assumptions in Section~\ref{s:simulation}. Some concluding remarks are provided in Section~\ref{s:discuss}.

\section{Problem Formulation}
\label{s:problem}
\subsection{Notation}
\label{s:problem:subs:notation}
  Suppose there are $n$ subjects entering the study at $E_i$, $i=1,\dots,n$, within the accrual period $[0,R]$. Let $T_i$ denote a univariate event time of interest, and $A_i$ indicate treatment assignment, e.g. using $0$ for the control group and $1$ for the treatment group. Given the treatment $A_i=a$, $a\in\{1,0\}$, event time $T_i$ follows a survival function $S_a(s)=P(T_i > s\mid A_i=a)$. To consider censoring time $C_i$, the observed follow-up time and event indicator are $\{Y_i(t)=min(T_i,C_i, t-E_i)\}$ and $\delta_i(t)=I(Y_i(t)=T_i)$, where $t$ is a stopping time for tests.  Alternatively, observed event times can also be written in a counting process form, i.e., $N_i(t,s)=I(T_i\leq s, \delta_i(t)=1)$. Note that $t$ and $E_i$ follow the chronological time scale, while $s$ in all the functions, $T_i$ and $C_i$ are in a follow-up time scale starting from the accrual time $E_i$. The censoring might differ between the two treatment arms so that the survival functions of censoring are $S_{ca}(s)=P(C_i > s\mid A_i=a)$, $a=0,1$. 

In this report, we allow both the control and the treatment group to follow piecewise exponential distributions with a general form: 
\begin{equation}\label{equ:survival}
S_a(s)= \exp\left[-\sum_{q=1}^{Q}\lambda_{aq} \max\left\{0,\min(\epsilon_q-\epsilon_{q-1},s-\epsilon_{q-1})\right\}\right],\quad a=1,0.
\end{equation} Note that $0=\epsilon_0<\epsilon_1<\dots<\epsilon_{Q}=\infty$ are the splitting points where the hazard changes, and $\lambda_{aq}$ is the hazard in the interval $[\epsilon_{q-1},\epsilon_q)$, $q = 1,\dots, Q$. The flexibility of including multiple pieces enables an accurate approximation of any survival curves. The corresponding density function is $f_a(s)=S_a(s)\sum_{q=1}^QI(s\in [\epsilon_{q-1},\epsilon_q))\lambda_{aq}$. The hazard ratios between the two treatment arms are $\bm\Theta=\{\theta_q=\lambda_{1q}/\lambda_{0q},q=1,\dots,Q\}$. The hazard ratios can describe all different changing pattern of the treatment effects, e.g., constant hazard ratio or PH cases with $\theta_q$ identical for $\forall q$, and delayed or increasing effects with $0<\theta_{Q}<\dots<\theta_1\leq1$ etc. For delayed treatment effect, a simple NPH case is given in Appendix~B (\ref{equ:survival_2p}) and (\ref{equ:density_2p}) with only the two-piece exponential distribution considered: hazards are $\lambda_{11}=\lambda_{01}=\lambda$ and $\theta_1=1$ for $[0,\epsilon)$, $\lambda_{12}=\theta_2\lambda$ and $\lambda_{02}=\lambda$ for $[\epsilon,\infty)$. In this simple case, the null hypothesis ($H_0$) has $\theta_2=1$ and the alternative hypothesis ($H_1$) has $\theta_2=\theta$ with some $\theta<1$. Or more broadly speaking, the null hypothesis $H_0$ is always $\bm\Theta=\bm\Theta_0$ with all elements $\theta_q=1$, and the alternative one $H_1$ could embrace any predefined $\bm \Theta=\bm\Theta_1$ with at least one of elements $\theta_q\not=1$. Additionally, we assume uniform accrual and administrative right censoring at time $\tau$ following the proposal of Hasegawa \cite{hasegawa2014sample} in Section~\ref{s:solution} when developing our solutions. However, the robustness of the proposed approaches towards the assumption violations can be seen in Section~\ref{s:simulation}, and easy extensions on the proposed method could be incorporated to accommodate more complicated scenarios in future studies. For example, one could implement the method from Luo et al \cite{luo2019design} for all types of piece-wise exponential distributions under various censoring and accrual processes. 

\subsection{Weighted log-rank test}
\label{s:problem:subs:WLRT}
Suppose the treatment-specific at-risk proportions are $R_a(t,s)=\frac{1}{n}\sum_{i=1}^n I(Y_i(t)\geq s, A_i=a)$, and the total at-risk proportion is $R(t,s)=R_1(t,s)+R_0(t,s)$. The standardized Fleming-Harrington class weighted log-rank test statistic (WLRT) stopped at time $t$ is given by 
\begin{equation}\label{equ:wlrt}
\mathcal{G}_{\rho,\gamma}(t)=\frac{\sum_{i=1}^n\int_0^t w_{\rho,\gamma}(s)[A_i-\frac{R_1(t,s)}{R(t,s)}]N_i(t,ds)}{\sqrt{\sum_{i=1}^n\int_0^t w_{\rho,\gamma}^2(s)\frac{R_1(t,s)R_0(t,s)}{R(t,s)^2}N_i(t,ds)}},
\end{equation} with a Fleming-Harrington weight $w_{\rho,\gamma}(s)=S(s^-)^\rho\{1-S(s^-)\}^\gamma$. We denote the numerator of (\ref{equ:wlrt}) as $G_{\rho,\gamma}(t)$, and its asymptotic variance $V(G_{\rho,\gamma}(t))$ can be estimated via the denominator:
\begin{equation}\label{equ:var_est}
\widehat{V}(G_{\rho,\gamma}(t))=\sum_{i=1}^n\int_0^t w_{\rho,\gamma}^2(s)\frac{R_1(t,s)R_0(t,s)}{R(t,s)^2}N_i(t,ds).
\end{equation} Thus formula in (\ref{equ:wlrt}) can also be written into and should satisfy the following equation $\mathcal{G}_{\rho,\gamma}(t)=G_{\rho,\gamma}(t)/\sqrt{\widehat{V}(G_{\rho,\gamma}(t))}=G_{\rho,\gamma}(t)/\sqrt{V(G_{\rho,\gamma}(t))}+op(1)$. 

Modifying the two parameters $\rho$ and $\gamma$ can adjust the shape of the weights, and thus the focus of the detection. For instance, $\rho=\gamma=0 $, the test statistic is reduced to a standard log-rank test (SLRT), which provides best power in the presence of PH; while $\rho=0$ and $\gamma=1$, it emphasizes more on late separations and thus provides better power for delayed effect. 

Sample size calculation for a confirmatory clinical trial is based on the predefined null and alternative hypotheses (denoted by $H_0$ and $H_1$). Letting $z_\alpha$ and $z_{1-\beta}$ be critical values of standard normal distribution, and $\Delta$ the effective difference between two arms (e.g. the constant log hazard ratio under PH),
The required number of events for SLRT has a closed-form expression \cite{schoenfeld1983sample}  given by
\begin{equation}
d=\frac{(z_\alpha+z_{1-\beta})^2}{p(1-p)\Delta^2}.
\end{equation} The sample size computation for WLRT, however, was established following a stochastic process approach suggested by Lakatos\cite{lakatos1988sample} and Hasegawa\cite{hasegawa2014sample}.  Suppose $b$ is the number of intervals at each time unit (month), and $J(t)=floor(bt)$ is the total number of time intervals at an equal length $[s_0=0,s_1,s_2,\dots,s_J=t]$, where $t$ represents a stopping time as aforementioned. There follows the mean estimator $\widetilde{E}_{sto,H}(G_{\rho,\gamma}(t))$ and variance/information estimator of the numerator in (\ref{equ:wlrt}):
\begin{equation}\label{equ:mean_sto}
\widetilde{E}_{sto,H}(G_{\rho,\gamma}(t))=\sum_{j=0}^{J(t)-1}D^*_{j,H}(t)w_{\rho,\gamma}^*(j)\left[\frac{\phi_j\theta_j}{1+\phi_j\theta_j}-\frac{\phi_j}{1+\phi_j}\right], H=H_1, H_0;
\end{equation}
\begin{equation}\label{equ:var_sto}
\widetilde{V}_{sto,H}(G_{\rho,\gamma}(t))=\sum_{j=0}^{J(t)-1}D^*_{j,\bm\Theta}(t)w_{\rho,\gamma}^{*2}(j)\frac{\phi_j}{(1+\phi_j)^2};
\end{equation}
where,
\begin{equation}\label{equ:compo_sto}
\begin{array}{l}
R_1^*(0)=p, R_0^*(0)=1-p,w_{\rho,\gamma   }^*(j)=S(s_j)^\rho[1-S(s_j)]^\gamma,\\
R_a^*(j+1)=R_a^*(j)\left[1-h_a^*(s_j)\frac{1}{b}-\frac{I(s_j>\tau-R)}{b(\tau-s_i)}    \right],  h_a^*(s_j)=\frac{f_a(s_j)}{S_a(s_j)},a=0,1,\\
\theta_j^*=\frac{h_1(s_j)}{h_0(s_j)}, \phi_j^*=\frac{R_1^*(s_j)}{R_0^*(s_j)},  D_{j,H}^*(t)=\left[\left\{h_0^*(s_j)R_0^*(j)+h_1^*(s_j)R_1^*(j)\right\}\frac{1}{b}\right]\min(\frac{t}{R},1).
\end{array}
\end{equation} Note that we set $H=H_1$ for the alternative hypothesis and  $H=H_0$ the null hypothesis in formulas (\ref{equ:mean_sto})-(\ref{equ:compo_sto}). It is also expected to see that $\widetilde{E}_{sto,H_0}(G_{\rho,\gamma}(t))=0$ in that $\theta_j=1$ for $\forall j$. In this report, we name the mean and variance in (\ref{equ:mean_sto})-(\ref{equ:var_sto}) be ``predicted'' values since none of the observation data are used here. Moreover, we use accent tilde to distinguish them from those estimated from observed events in (\ref{equ:var_est}). The difference between the prediction and estimation methods merely depends on whether the observed event times are used in the computation. Henceforth, prediction is usually used for early trial design purpose, but estimation is more often implemented when part of the data have been collected. The marginal survival function is the weighted average of two treatment arms: $S(s)=(1-p)S_0(s)+pS_1(s)$. In contrast to the proposal by Hasegawa \cite{hasegawa2014sample}, there is an extra multiplicative term $\min(\frac{t}{R},1)$ for the formulation of $D_{j,\bm\Theta_1}^*(t)$ in (\ref{equ:compo_sto}) to account for a special case that $t<R$, i.e., stopping before the end of accrual period.

Lakatos \cite{lakatos1988sample} suggested that $\mathcal{G}_{\rho,\gamma}(t)$ follows an asymptotically normal distribution with unit variance and mean approximated by
\begin{equation}\label{equ:mu_sto}
\widetilde{\mu}_{\rho,\gamma,H}(t)=\frac{\widetilde{E}_{sto,H}(G_{\rho,\gamma}(t))}{\sqrt{\widetilde{V}_{sto,H}(G_{\rho,\gamma}(t))}}. 
\end{equation} Fixing $t=\tau$ for tests at the end of the study, the required sample sizes in terms of the total number of subjects ($n$) and observed events ($d$) are
\begin{equation}
\begin{array}{l}
n=\left(\frac{z_\alpha+z_{1-\beta}}{ \widetilde{\mu}_{\rho,\gamma,H_1}(\tau)}\right)^2\\
d=nD_{H_1}^*(\tau),
\end{array}
\end{equation} where $D_{H_1}^*(\tau)=\sum_{j=0}^{J(\tau)-1}D^*_{j,H_1}(\tau)$ approximates the probability of observing an event from each subject under $H_1$.

\subsection{Maxcombo test}
\label{s:problem:subs:maxcombo}
In practice, the true survival curves or the hazard ratio between the treatment arms are usually unknown; moreover, the existence of delayed treatment effect and its severity can hardly be predicted in advance. To that end, Lee \cite{lee1996some} proposed a versatile max-combo test, taking the maximum of a combo of different WLRTs to cover various scenarios: PH case, NPH cases with early, middle and late effects, etc. The general form of a maxcombo test is  
\begin{equation}\label{equ:maxcombo}
\mathcal{G}_{max}(t)=\max\left(\mathcal{G}_{\rho_1,\gamma_1}(t), \mathcal{G}_{\rho_2,\gamma_2}(t),\dots,\mathcal{G}_{\rho_K,\gamma_K}(t)\right),
\end{equation} where $\mathcal{G}_{\rho_k,\gamma_k}(t)$ is one of the $K$ different Fleming-Harrington family WLRTs. Boundary calculation for a maxcombo test statistic is equivalent to finding the boundary value $ g(\tau)$ at the end of the study (time $\tau$) such that the type one error ($\alpha$) is under control: 
\begin{equation}\label{equ:alpha}
P(\mathcal{G}_{max}(\tau)<g(\tau)\mid H_0)=1-\alpha.
\end{equation} According to Lee \cite{lee2007versatility}, $ \bm{\mathcal{ G}}(t)=[\mathcal{G}_{\rho_1,\gamma_1}(t),\dots,\mathcal{G}_{\rho_K,\gamma_K}(t)]^\prime$ is (asymptotically) multivariate normal distributed with mean 0 and variance 1, and the correlation for two different WLRTs with $k_1\not=k_2$ is given by
\begin{equation}\label{equ:cor}
Cor(\mathcal{G}_{\rho_{k_1},\gamma_{k_1}}(t),\mathcal{G}_{\rho_{k_2},\gamma_{k_2}}(t))=\frac{Cov(G_{\rho_{k_1},\gamma_{k_1}}(t),G_{\rho_{k_2},\gamma_{k_2}}(t))}{\sqrt{V(G_{\rho_{k_1},\gamma_{k_1}}(t))V(G_{\rho_{k_2},\gamma_{k_2}}(t))}}.
\end{equation}

The variances can be obtained through either the data-driven estimation $\widehat{V}(G_{\rho,\gamma}(t))$ from (\ref{equ:var_est}) or the stochastic prediction $\widetilde{V}_{sto}(G_{\rho,\gamma}(t))$ from (\ref{equ:var_sto}). In a similar vein, covariance can be obtained following either prediction or estimation via the equation given by
\begin{equation}\label{equ:cov}
Cov(G_{\rho_{k_1},\gamma_{k_1}}(t),G_{\rho_{k_2},\gamma_{k_2}}(t))=V\left\{G_{\frac{\rho_{k_1}+\rho_{k_2}}{2},\frac{\gamma_{k_1}+\gamma_{k_2}}{2}}(t)\right\}.
\end{equation}

Alternatively, for the (piece-wise) exponential distributions, one can derive close-form expressions for their mean, variance, and covariance values according to their exact distribution functions, for which we denote as ``exact prediction'', and indexing with an ``exa'' for distinction. Please check the exactly predicted variances ($\widetilde{V}_{exa}$) and covariance ($\widetilde{Cov}_{exa}$) for a piece-wise exponential survival distribution in Appendix~B. The exact prediction method can largely alleviate the computational burden, though the closed-form solutions may not exist for complex survival curves. To that end, one might refer to numerical approximation by transforming the integration to a summation over many small intervals, which is quite similar to the proposed stochastic prediction method. 

Under $H_1$, the asymptotic mean for each WLRT can be approximated through (\ref{equ:mu_sto}), and thus the mean vector $\widetilde{\bm\mu}(t)=[\widetilde{\mu}_{\rho_1,\gamma_1,H_1}(t),\dots,\widetilde{\mu}_{\rho_K,\gamma_K,H_1}(t)]^\prime$. The approximate asymptotic distribution of the test statistics $\bm {\mathcal{G}}(t)$ is multivariate normal with mean $\sqrt{n}\widetilde{\bm\mu}(t)$ and the covariance/correlation matrix can be obtained via the prediction methods proposed for the boundary calculation in (\ref{equ:cor}). Note that we do not consider using the estimation approach for sample size calculation, since sample size calculation is usually decided before starting the trial in group sequential designs. The sample size will be obtained through solving the function below so that the type II error equals $\beta$: 
\begin{equation}\label{equ:beta}
P(\mathcal{G}_{max}(\tau)<g(\tau)\mid H_1)=\beta. 
\end{equation}

\subsection{Group Sequential Design for Maxcombo tests}
\label{s:problem:subs:IA-MC}
Practitioners often employ interim analyses or group sequential designs in clinical trials.  They will save time and budgets by stopping a trial early when there is sufficient statistical evidence to terminate the study:  futility, unexpected side effects, and significant treatment effect. There were extensive discussions about introducing maxcombo tests to group sequential design on the FDA workshop at Duke-Margolis Health Policy Center in 2018 \cite{fdaworkshop}. Plenty of simulations have shown that maxcombo could potentially improve the robustness when NPH exists. However, there are mainly two problems that hinder the implementation of GS-MC: 1)  how to compute the boundaries at each stage to control the type I error; 2) how to compute the sample size.  Simulations can solve both two problems, but the computational burden could be considerable. To avoid the tedious simulations, we propose a design procedure that can control the type I and accurately predicted the required sample size by approximating the asymptotic distribution of all the test statistics across different stopping points. 

\section{Proposed solutions}
\label{s:solution}
%For simplicity, we only include two test statistics $G_{\rho_1,\gamma_1}(t)$ and $G_{\rho_2,\gamma_2}(t)$. The maxcombo of the two tests is $G_{max}(t)=\max(G_{\rho_1,\gamma_1}(t),G_{\rho_2,\gamma_2}(t))$. Suppose there are only one interim stage at time $t_1$ before the final stage at $\tau$, we have 4 test statistics $G_{\rho_1,\gamma_1}(t_1)$, $G_{\rho_2,\gamma_2}(t_1)$, $G_{\rho_1,\gamma_1}(\tau)$, and $G_{\rho_2,\gamma_2}(\tau)$, where $t_1<\tau$. The following subsections aim to calculate their correlation matrices and means under both hypotheses ($H_0$ and $H_1$). Method proposed here can be extended to accommodate more than two tests and multiple interim stages.

\subsection{Correlation matrix approximation}
\label{sub:correlation}
The correlation matrix of the test statistics requires 3 different types of correlation values. The first type is the within-stage correlation between different tests, e.g. $Cor(\mathcal{G}_{\rho_{k_1},\gamma_{k_1}}(t),\mathcal{G}_{\rho_{k_2},\gamma_{k_2}}(t))$, which can be computed following equation (\ref{equ:cor}). The second type is within-test correlation between two stopping time points, or $Cor(\mathcal{G}_{\rho,\gamma}(t_{m_1}),\mathcal{G}_{\rho,\gamma}(t_{m_2}))$ for $0<t_{m_1}<t_{m_2}\leq\tau$ and $m_1<m_2$, computed from
\begin{equation}\label{equ:cor_t}
Cor(\mathcal{G}_{\rho,\gamma}(t_{m_1}),\mathcal{G}_{\rho,\gamma}(t_{m_2}))=\sqrt{\frac{V(G_{\rho,\gamma}(t_{m_1}))}{V(G_{\rho,\gamma}(t_{m_2}))}}.
\end{equation} The information fraction $IF_{\rho,\gamma}(t_{m_1},t_{m_2})={V(G_{\rho,\gamma}(t_{m_1}))}/{V(G_{\rho,\gamma}(t_{m_2}))}$ under the square root of (\ref{equ:cor_t}) was used to decide stopping times in group sequential designs \cite{hasegawa2016group}. Note that (\ref{equ:cor_t}) asymptotically holds only under $H_0$, when the independent increment property $Cov(G_{\rho,\gamma}(t_{m_1}),G_{\rho,\gamma}(t_{m_2}))=V(G_{\rho,\gamma}(t_{m_1}))$ is asymptotically true \cite{tsiatis1981asymptotic}. Although not strictly satisfied under $H_1$, the independent increment property and thus the equation (\ref{equ:cor_t}) almost hold numerically when the difference between two treatment arms are not considerable (under the so-called ``local alternatives") or when the events are not too frequent, according to Example~1 in Luo et al \cite{luo2019design} and our simulations in Section~\ref{s:simulation}. Note that the variance value for both within-test and within-stage correlations can be obtained via both prediction and estimation approaches. 

The third type includes correlations across different time points and test types. We propose a simple calculation based on the first two types of correlations by introducing the Theorem~1, which is also proved in Appendix~\ref{app:sec:proof}. 

\begin{theorem}
	If random variables $X_1$, $X_2$ and $X_3$ have mean 0, variance 1, satisfying $X_3=\phi X_2+M$, $M\perp (X_1,X_2)$ with $E(M)=0$, and $\phi$ is a constant value, then the equality $cor(X_1,X_3)= cor(X_1,X_2)cor(X_2,X_3)$ holds. 
\end{theorem}

In particular, let $X_1=\mathcal{G}_{\rho_{k_1},\gamma_{k_1}}(t_{m_1})-E[\mathcal{G}_{\rho_{k_1},\gamma_{k_1}}(t_{m_1})]$ and $X_2=\mathcal{G}_{\rho_{k_2},\gamma_{k_2}}(t_{m_1})-E[\mathcal{G}_{\rho_{k_2},\gamma_{k_2}}(t_{m_1})]$, then it holds that $X_3=\mathcal{G}_{\rho_{k_2},\gamma_{k_2}}(t_{m_2})-E[\mathcal{G}_{\rho_{k_2},\gamma_{k_2}}(t_{m_2})]=\phi X_2+M$, where $\phi=\sqrt{IF_{\rho_{k_2},\gamma_{k_2}}(t_{m_1},t_{m_2})}$ and
$M=[G_{\rho_{k_2},\gamma_{k_2}}(t_{m_2})-G_{\rho_{k_2},\gamma_{k_2}}(t_{m_1})-E\{G_{\rho_{k_2},\gamma_{k_2}}(t_{m_2})-G_{\rho_{k_2},\gamma_{k_2}}(t_{m_1})\}]/\sqrt{V(G_{\rho_{k_2},\gamma_{k_2}}(t_{m_2}))}$. Note that under the $H_0$ we have $E(G_{\rho,\gamma}(t))=0$. Similar to the comments for the second-type correlation, $M\perp (X_1,X_2)$ is asymptotically correct following the asymptotic independent increment property \cite{tsiatis1981asymptotic} under the $H_0$, and only approximately true under $H_1$ when the difference between two arms and the event hazards are limited within some practical range. In the Section~\ref{s:simulation}, we conducted extensive simulations to explore how well the this approximation is in various scenarios and in the presence of multiple assumption violations (Table~\ref{tab:corr}, Web Tables~3-6). Following Theorem~1, we have the third-type correlation given by
\begin{equation}\label{equ:equal}
\begin{array}{l}
Cor(\mathcal{G}_{\rho_{k_1},\gamma_{k_1}}(t_{m_1}),\mathcal{G}_{\rho_{k_2},\gamma_{k_2}}(t_{m_2}))=\\
\qquad\qquad Cor(\mathcal{G}_{\rho_{k_1},\gamma_{k_1}}(t_{m_1}),G_{\rho_{k_2},\gamma_{k_2}}(t_{m_1}))Cor(\mathcal{G}_{\rho_{k_2},\gamma_{k_2}}(t_{m_1}),\mathcal{G}_{\rho_{k_2},\gamma_{k_2}}(t_{m_2})). 
\end{array}
\end{equation}
With all the 3 different types of correlations calculated using either distribution-based prediction or data-driven estimation, the two sets of correlation matrices are obtained under $H_0$ and $H_1$.

\subsection{Type I error: boundaries}
\label{sub:type1error}
We introduce boundary vector $\bm g=[g(t_{1}),\dots,g(t_{M})]^\prime$ for M stages including a final stage and $M-1$ interim stages. To control the type I error at each stage, we employ a monotone increasing error spending function $\alpha(\nu)$ with $\nu\in[0,1]$, $\alpha(0)=0$ and $\alpha(1)=\alpha$ \cite{gordon1983discrete}. Suppose that we monitor the information fractions at times $0=t_0<t_1<\dots<t_M=\tau$ satisfying $IF_{\rho_{k},\gamma_{k}}(t_{m},\tau)=\nu_m$, where $0=\nu_0<\nu_1<\dots<\nu_M=1$ are pre-defined, and $w_{\rho_k,\gamma_k}$ indicates the $kth$ WLRTs in the combo to monitor stopping times. Error spending funcitonn $\alpha(v)$ controls type I error spent at each stage via step-wise equations for stage $m=1,\dots,M$,
\begin{equation}
\begin{array}{l}
P(\mathcal{G}_{max}(t_{j})\leq g(t_{j}),j=1,\dots,m-1,\mathcal{G}_{max}(t_{m})>g(t_{m})\mid H_0)=\alpha(\nu_m)-\alpha(\nu_{m-1}).
\end{array}
\end{equation} The boundary values $\bm g$ can be obtained via solving the multivariate normal distribution with mean $0$, and variance matrix $\bm \Sigma_0$ with all the diagonal entries to be 1, and off-diagonal correlation entries computed following Subsection~\ref{sub:correlation}. 

\subsection{Power: sample size}
The sample size calculation is based on the asymptotic distribution of the multivariate normal distribution under $H_1$. With all the boundaries decided in Subsection~\ref{sub:type1error}, the sample sizes ($n$) can be solved such that we have
\begin{equation}\label{equ:beta2}
\begin{array}{l}
P(\mathcal{G}_{max}(t_{m})\leq g(t_{m}),m=1,\dots,M\mid H_1)=\beta. 
\end{array}
\end{equation} In group sequential trials, sample size calculation usually precedes the trial. we will obtain the sample size according to predicted mean $\sqrt{n}[\widetilde{\bm\mu}(t_{1})^\prime,\dots,\widetilde{\bm\mu}(t_M)^\prime]^\prime$ and variance matrix $\widetilde{\bm\Sigma}_1$ of $\bm{ \mathcal{G}}=[\bm{\mathcal{ G}}(t_1)^\prime,\bm {\mathcal{G}}(t_2)^\prime,\dots,\bm{ \mathcal{G}}(t_M)^\prime]^\prime$, where $\bm{\mathcal{G}}(t_m)^\prime=[\mathcal{G}_{\rho_1,\gamma_1}(t_m),\dots,\mathcal{G}_{\rho_K,\gamma_K}(t_m)]$. The required event count is $d=nD_{H_1}^*(\tau)$. 

\subsection{A complete design}
\begin{figure}[h!]
	\centerline{\includegraphics[width=\columnwidth]{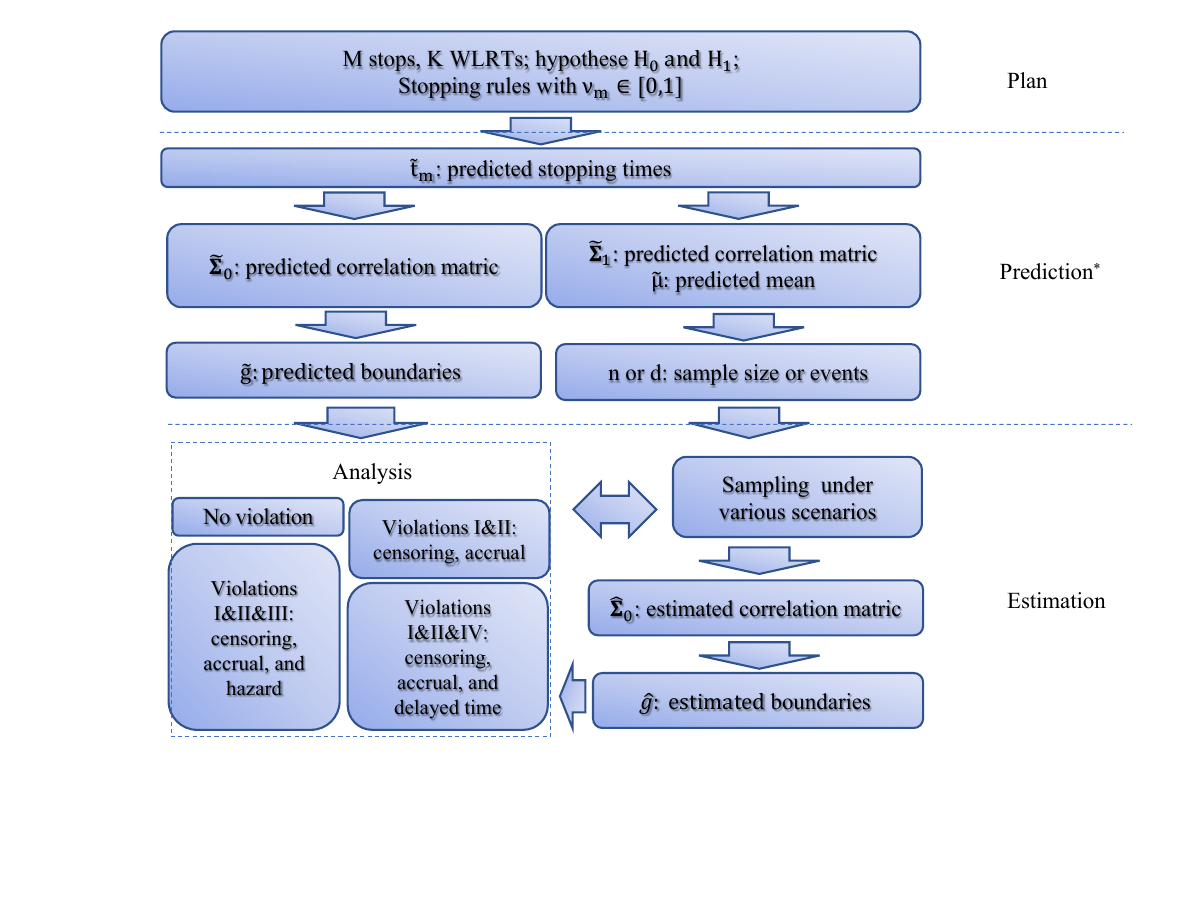}}
	\caption{A flow-chart to describe the procedure of the proposed simulation-free GS-MC design. The superscript ``*''  indicates that the correlation matrices can be predicted using both the stochastic and exact approaches, while the mean ($\widetilde{\mu}$)  is predicted using stochastic approach to enjoy a more precise approximation as the at-risk proportions change by time. The ``analysis'' stage is when we conducted the maxcombo tests under various scenarios with or without assumption violations, and the output of this step is the decision to reject or accept $H_0$, which are summarized as type I error and power in simulations. The arrows between the blocks indicate directions of the information flow. }
	\label{fig:1}
\end{figure}

The complete design is summarized in Figure~\ref{fig:1}. First, one will need to decide the plan by defining the hypotheses $H_0$ and $H_1$, their survival functions, $K$ WLRTs within the combo, $M-1$ interim stages, and the corresponding stopping rule $\nu_m$. The stopping rule is dependent on the information fraction of one of the WLRTs in the combo, i.e., stopping at $IF_{\gamma_{k},\rho_k}(t_m)=\nu_m$ for $mth$ test if the previous $m-1$ stages fail to reject $H_0$.  For instance, if using surrogate information fraction\cite{hasegawa2016group}, it would base on the event counts. The correlation matrices can be obtained using either the prediction approach or the estimation approach. Prediction can be done using the stochastic method and the exact method, whereby the former would apply to all kinds of survival functions and also consider the changing at-risk proportion of treatment group, while the latter treats this proportion a constant value (denoted to be p) in the formulas given in Appendix~B. The estimation approach is entirely data-driven, following equations (\ref{equ:var_est}) and (\ref{equ:cov}). 

The stopping times are predicted according to the predicted information fractions, and thus we obtain the resulting distributions of the multivariate test statistics under both hypotheses. As follows, the boundaries at each stage and sample size can be predicted as well. Moreover, once we start the trial, data are collected, and the correlation matrix can be estimated following (\ref{equ:var_est}). Therefore, instead of using the predicted correlations, one can also estimate the correlation matrices for boundary and sample size calculation, in order to ensure that the type I error can still be controlled when the assumed survival distributions, censoring process, or accrual procedure are violated in practice. 
\section{Simulation Studies}
\label{s:simulation}
We used the two-piece exponential provided in Appendix~B as an example to demonstrate and compare the performance of our proposed approaches. The event times from the control arm is following an exponential distribution with rate $\lambda=log(2)/6$ (median survival: 6 months), while the event times from the treatment arm were generated following a two-piece exponential with its hazard changing from $\lambda$ to $\theta \lambda$ at after $\epsilon=2$ months of follow-up. When $\theta=1$, the two-piece exponential is reduced to an exponential distribution identical to the control group. To strictly follow the piece-wise exponential distribution suggested in (\ref{equ:survival}) with $Q=2$, we let $\bm\Theta_0=\{1,1\}$ for $H_0$ and $\bm\Theta_1=\{1,\theta\}$ for $H_1$, where $\theta\in(0,1)$ and $\epsilon_1=\epsilon$, which is equivalent to (\ref{equ:survival_2p}). Simulations under different scenarios were carried out to evaluate the proposed methods for a reasonable range of post-delay treatment effects ($\theta\in[0.5,0.7]$). To begin, we generated data following uniform accrual within time interval $[0, R]$,  where $R=14$; and with a probability $p=0.5$, the subjects were randomly assigned to the treatment arm. All the studies were expected to end at $\tau=18$. We included two log-rank test statistics for a maxcombo test $\mathcal{G}_{max}(t)=\max(\mathcal{G}_{0,0}(t),\mathcal{G}_{0,1}(t))$, and one interim stopping stage. Thus we have $M=2$ and $K=2$. Note that $\mathcal{G}_{0,1}(t)$ tends to be more powerful than $\mathcal{G}_{0,0}(t)$ in the presence of a delayed treatment effect.  In practice, however, the existence of such a delayed effect is generally unknown. Moreover, its severity (in terms of $\theta$ and $\epsilon$) can hardly be predicted, thus incorporating more WLRTs can potentially provide better robustness. In the following simulations, we only focus on one-sided tests, with their type I error controlled at level $\alpha=0.25$ and sample size targeting power $1-\beta=0.9$, respectively. For each simulation study, we generated 200,000 datasets for type I error evaluation, and 50,000 for power estimation.

The stopping times were decided according to the information fraction of the SLRT $\mathcal{G}_{0,0}$, or namely the surrogate information fraction in Hasegawa \cite{hasegawa2016group}. We stopped for an interim analysis at $t_{int}$ when $0.6d$ events were observed, and terminated the study for a final analysis at $t_{fin}$ when $d$ events were observed.  In other words, we let $\nu_1=0.6$ and $\nu_2=1$. Note that $d$ is the total number of events we need and will be predicted once the stopping times are predicted. In particular, the stopping times ($t_{int}$ and $t_{fin}$) can be decided by solving $D^*_{H}(\widetilde{t}_{int})=0.6D_{H_1}^*(\tau)$ and $D^*_{H}(\widetilde{t}_{fin})=D_{H_1}^*(\tau)$, with $H=H_0$ when the null hypothesis is true, and $H=H_1$ otherwise. Or in other more general cases when monitoring any WLRT with respect to its information fraction, one would predict the stopping times by solving $\widetilde{V}_{sto,H}(G_{\rho,\gamma}(\widetilde{t}_{int}))=0.6\widetilde{V}_{sto,H_1}(G_{\rho,\gamma}(\tau))$ and $\widetilde{V}_{sto,H}(G_{\rho,\gamma}(\widetilde{t}_{fin}))=\widetilde{V}_{sto,H_1}(G_{\rho,\gamma}(\tau))$. Note that the two sets of stopping times can differ under different hypotheses, and consequently the predicted correlation matrices ($\widetilde{\bm \Sigma}_0$ and $\widetilde{\bm \Sigma}_1$) are different too. The mean of the test statistics under $H_0$ is $\bm0_4$, but is $[\widetilde{\bm\mu}(\widetilde{t}_{int})^\prime,\widetilde{\bm\mu}(\widetilde{t}_{fin})^\prime]^\prime$ under $H_1$. We obtained the predicted boundaries $\widetilde{\bm g}$ and sample sizes $d$ or $n$ based on the predicted mean and correlation matrices. We can also predict the stopping times and subsequently the correlation matrices using the exact-prediction method given in Appendix~B. Alternatively, the boundaries $\widehat{g}(t)$ can be updated according to the data by calculating the estimated correlation matrix $\widehat{\bm\Sigma}_0$ following (\ref{equ:var_est}) and (\ref{equ:cov}), which is expected to be more accurate than the prediction methods in the presence of violations in the distributional assumptions of the survival, censoring and accrual processes. We used R package mvtnorm \cite{mvtnorm} for the boundary calculation.  Since it is seed-dependent, we each time generated 5 replicates, and keep the median of them as output value. All the prediction and estimation methods proposed in this report have been established in an R package on Github (lilywang1988/GSMC). 

First, we tested various post-delay hazard ratios with only administrative censoring and correctly specified survival functions, in consistence with the assumptions given in Hasegawa \cite{hasegawa2014sample}. All results were summarized in Table~\ref{tab:default} , where  GS-WLRT denotes group sequential design with a WLRT $\mathcal{G}_{0,1}(t)$, GS-SLRT denotes group sequential design with SLRT $\mathcal{G}_{0,0}(t)$, and GS-MC denotes group sequential design with a maxcombo test of $\mathcal{G}_{0,1}(t)$ and $\mathcal{G}_{0,0}(t)$. Note that since all the tests were in group sequential designs, we eliminated prefix `GS'' in the summary tables. In GS-MC, we consider the naive method with boundaries that are identical to the other two univariate tests GS-WLRT and GS-SLRT, which are stopped according to the observed events (namely, the surrogate information fraction of SLRT according to Hasegawa\cite{hasegawa2016group}). To produce fair comparisons,  we computed sample sizes for GS-MC using proposed prediction methods (the stochastic and exact methods), whose results turned out to be identical. The boundaries for the other three GS-MC methods were computed from the prediction (stochastic or exact), and the estimation approaches as well. As expected, in the presence of a delayed treatment effect, GS-SLRT provided a controlled type I error for the correct boundary specification under $H_0$, but the powers were much lower than the rest because it did not consider the delayed treatment effect well. On the other hand, GS-WLRT provides a much higher power than GS-SLRT, but the type I error ($0.0259-0.0268$) was obviously above the nominal $0.025$, which is because the surrogate information fraction did not reflect the true correlation between the two stopping times for WLRT, or in other words, the boundaries for GS-SLRT were not appropriate for GS-WLRT. To that end, it was suggested to monitor the true information fractions of WLRT instead of using the surrogate information fraction by Hasegawa \cite{hasegawa2016group}.  The naive GS-MC appeared to enjoy a higher power than WLRT, but its type I error ($\approx0.04$) was also not controlled under the nominal $0.025$. This is another example that the event count ratio (surrogate information fraction) did not reflect the correlation between the two maxcombo tests at the interim and final stages. 

There did not seem to exist much difference comparing the performance among the proposed approaches, though the estimation approach had slightly better power than the two predicted methods with their type I error controlled similarly. The stochastic prediction does not limit its survival function to piece-wise exponential distributions. \iffalse Its flexibility seems not to improve its performance in comparison with the exact prediction approach, possibly because its prediction on the stopping times are not as accurate as those using exact prediction.\fi  All the three proposed approaches performed much better than the naive method in controlling the type I error. By increasing the post-delay separation up to $\theta=0.5$, the type I error increased slightly above $0.025$, and the power decreased. There could be mainly two explanations: 1) the accrual sample size decreased from $n=927$ ($d=597$) to $n=274$ ($d=166$) when the post-delay hazard ratio was decreased from $\theta=0.7$ to $\theta=0.5$,  thus the tails of the distribution for the test statistic became heavier; 2) the increasing treatment effect caused a more serious violation of the independent increment assumption, and thus damaged the approximation accuracy. 

%%%%  table 1  %%%%
%\begin{sidewaystable}
\begin{table}\centering
	
	\caption{The rejection probabilities under the null hypothesis denoted by $H_0$ (type I error) and under the alternative hypothesis $H_1$ (power) when \emph{censoring, accrual, and survival functions are correctly specified}. Prefixes ``GS''  standing for ``group-sequential'' are eliminated here for simplicity. Sample sizes were decided according to GS-MC, and both prediction approaches provided identical sample sizes. Note that among the proposed GS-MC methods, the predicted powers are $0.3692$, $0.3630$, $0.3543$ for $\theta\in\{0.7,0.6,0.5\}$ at the interim stage respectively, and all are $0.9$ with two stages combined. }
	\label{tab:default}
	\begin{tabular}{lllcclcclccl}\hline
		\phantom{}&{}&\phantom{}&\multicolumn{2}{c}{ $\theta=0.7$} &
		&\multicolumn{2}{c}{ $\theta=0.6$} &\phantom{}&\multicolumn{2}{c}{ $\theta=0.5$} &\phantom{}\\ [-1pt]
		\cline{4-5} \cline{7-8} \cline{10-11}\\[-10pt]
		&Test&Stage&$H_0$&$H_1$& &$H_0$&$H_1$& & $H_0$&$H_1$& \\ \hline
		&WLRT& combined           &0.0260&0.9143& &0.0262&0.9140& &0.0269&0.9137&\\
		&    & interim            &0.0051&0.4082& &0.0052&0.4039& &0.0055&0.3983&\\
		&SLRT& combined           &0.0251&0.8272& &0.0252&0.8206& &0.0246&0.8103&\\
		&    & interim            &0.0051&0.2487& &0.0049&0.2355& &0.0050&0.2283&\\
		&MC (naive)&combined   &0.0384&0.9279& &0.0388&0.9273& &0.0390&0.9258&\\
		&    & interim            &0.0082&0.4346& &0.0082&0.4282& &0.0085&0.4213&\\
		&MC (pred-sto)&combined&0.0251&0.8970& &0.0253&0.8962& &0.0255&0.8948&\\
		&                &interim &0.0050&0.3691& &0.0051&0.3620& &0.0054&0.3579&\\
		&MC (pred-exa)&    combined &0.0252&0.8972& &0.0253&0.8963& &0.0255&0.8950&\\
		&           &     interim &0.0050&0.3693& &0.0051&0.3621& &0.0054&0.3582&\\
		&MC (est)&    combined &0.0252&0.8972& &0.0253&0.8966& &0.0255&0.8952\\
		&           &     interim &0.0050&0.3693& &0.0051&0.3622& &0.0054&0.3581&\\
		&size&  &n= 927&d=597&	&n=475&d=297&&n=274&d=166&\\
		\hline
	\end{tabular}	 
	\vspace{5pt}
\end{table}
%\end{sidewaystable}
%%%%  end table 1  %%%%

Then we tested different special cases with some of the assumptions, e.g., uniform accrual, administrative censoring, correctly specified survival hazards, and delayed time being violated. For violation I, the accrual is no longer uniform with a monthly accrual rate $n/14$ subjects per month, but instead is $n/70$, $2n/70$, $3n/70$, $4n/70$ for the first 4 months, and then $6n/70$ for rest 10 months. In the presence of violation I, still there are $n$ subjects enrolled by the end of the accrual period $R=14$, but the accrual rate is increasing for the first four months before getting stabilized at a constant rate. For violation II, censoring is not limited to a shared administrative censoring but can differ between treatment arms. In particular, we generated censoring time following exponential distributions with a yearly censoring proportion $20\%$ for the treatment group and $10\%$ for the control group. For violation III, the true median survival time for the control group is 12 months, other than the presumed six months. For violation IV, the separation occurs at six months after the enrollment, instead of the predefined two months. The results under various post-delay treatment hazard ratios with violations I and II (I\&II) were summarized in Table~\ref{tab:violation1}, and those with an additional violation III (I\&II\&III) or violation IV (I\&II\&IV) were in Web Tables~1-2. Their corresponding correlation matrices can be found in Web Tables~3-6, and the subset results of $\theta=0.6$ in Table~\ref{tab:corr}.  

According to Table~\ref{tab:violation1}, it turned out that all the three proposed approaches were quite robust to misspecifications of the accrual process and the censoring mechanism (violations I\&II).  The average stopping times ($t_{int}$, $t_{fin}$) were larger than those from Table~\ref{tab:default}, to collect enough events (information) in front of the additional censoring. The detection power was even better, possibly due to a long waiting time before ending the study enabled us to collect more events after the start of the separation. Similar to the results without any violations in Table~\ref{tab:default}, the estimation approach has slightly better power than the two prediction methods, with all their type I error controlled similarly. Note that since the subjects were enrolled slower than those with uniform enrollment in Table~\ref{tab:default} at the early stage, the power was generally lower at the interim stage.

Other additional violations were considered in the Web Tables~1-2. When there existed another violation that the event hazard $\lambda=\log(2)/12$ was wrongly specified to be $\lambda=\log(2)/6$ (I\&II\&III), the waiting time before observing enough events would be longer than that without violation III in Tables~\ref{tab:default}-\ref{tab:violation1}, and thus producing a higher power for delayed separation according to Web Table~5. Or when the delayed effect time $\epsilon=6$ was misspecified to be $\epsilon=2$ in addition to the violations in censoring and accrual (I\&II\&IV), the power would become much lower than the expected value according to Web Table~6, since the required sample sizes to achieve the nominal power was largely underestimated. Among all the combinations of violations we tested, the type I error was not affected much, while the power was obviously affected depending on the degree of violations of the model assumptions. 

To scrutinize whether the correlation matrices were accurately approximated and how they were affected by the violations, we compared correlations computed using the prediction and estimation approaches with or without violations I\&II for $\theta=0.6$ in Table~\ref{tab:corr}, and all other cases in Web Tables~3-6. The mean correlations of the simulated datasets were treated as gold standards. When there was no assumption violation, none of the correlations had more than $5\%$ difference from the gold standard (Table~\ref{tab:corr} and Web Table~3). Only the predicted $\widetilde{cor}(\mathcal{G}_{0,1}(t_{int}),\mathcal{G}_{0,1}(t_{fin}))$ and $\widetilde{cor}(\mathcal{G}_{0,0}(t_{int}),\mathcal{G}_{0,1}(t_{fin}))$ were over $5\%$ different from the gold standard correlations under $H_1$  (Table~\ref{tab:corr} and Web Table~4), possibly because the violations in censoring and accrual mechanisms affect the prediction of the stopping times, the predicted correlations of $G_{0,1}(t)$ are not correctly reflecting the true correlations between stopping times. Note that $cor(\mathcal{G}_{0,0}(t_{int}),\mathcal{G}_{0,1}(t_{fin}))$ is approximated by the product of $cor(\mathcal{G}_{0,1}(t_{int}),\mathcal{G}_{0,1}(t_{fin}))$ and $cor(\mathcal{G}_{0,0}(t_{int}),\mathcal{G}_{0,1}(t_{int}))$ following (\ref{equ:equal}). In the presence of violations I\&II\&III, the predicted correlations for $cor(\mathcal{G}_{0,1}(t_{int}),\mathcal{G}_{0,1}(t_{fin}))$ and $cor(\mathcal{G}_{0,0}(t_{int}),\mathcal{G}_{0,1}(t_{fin}))$ would suffer from a severe bias with over $25\%$ difference from the gold standard values. Misspecification of the delayed time, on the contrary, did not impact the correlation matrices much. The estimation method provides the most accurate correlation approximations, and the correlations predicted via the stochastic method is slightly more accurate than those via the exact prediction method. The type I error, however, seems not being affected much by the correlation matrices. We checked the boundaries predicted and estimated under different violation combinations and found that their changes were extremely small ($<0.003$ in magnitudes), implying that slight bias in the correlations did not exert considerable influence on the boundary calculation under $H_0$. %  It is entirely possible that when there are more candidate tests included and more stopping stages in the GS-MC study, these violations might lead to severer biases in the correlation matrix computation.   

%%%%  table 2  %%%%
%\begin{sidewaystable}
\begin{table}\centering
	\caption{The rejection probabilities under the null hypothesis denoted by $H_0$ (type I error) and under the alternative hypothesis $H_1$ (power) \emph{when censoring and accrual are misspecified}. Prefixes ``GS''  standing for ``group-sequential'' are eliminated here for simplicity. }
	\label{tab:violation1}
	\begin{tabular}{lllcclcclccl}\hline
		\phantom{}&{}&\phantom{}&\multicolumn{2}{c}{ $\theta=0.7$} &
		&\multicolumn{2}{c}{ $\theta=0.6$} &\phantom{}&\multicolumn{2}{c}{ $\theta=0.5$} &\phantom{}\\ [-1pt]
		\cline{4-5} \cline{7-8} \cline{10-11}\\[-10pt]
		&Test&Stage&$H_0$&$H_1$& &$H_0$&$H_1$& & $H_0$&$H_1$& \\ \hline
		&WLRT& combined           &0.0259&0.9172& &0.0261&0.9159& &0.0268&0.9135&\\
		&    & interim            &0.0052&0.3747& &0.0053&0.3663& &0.0052&0.3569&\\
		&SLRT& combined           &0.0248&0.8325& &0.0248&0.8224& &0.0250&0.8128&\\
		&    & interim            &0.0053&0.2156& &0.0051&0.2002& &0.0051&0.1917&\\
		&MC (naive)&combined      &0.0379&0.9293& &0.0383&0.9279& &0.0391&0.9252&\\
		&    & interim            &0.0085&0.3996& &0.0084&0.3883& &0.0084&0.3782&\\
		&MC (pred-sto)&combined   &0.0246&0.9010& &0.0247&0.8976& &0.0256&0.8953&\\
		&    &interim             &0.0052&0.3339& &0.0052&0.3255& &0.0051&0.3145&\\
		&MC (pred-exa)&    combined &0.0246&0.9012& &0.0248&0.8976& &0.0256&0.8954&\\
		&           &     interim  &0.0053&0.3342& &0.0052&0.3258& &0.0051&0.3147&\\
		&MC (est)&    combined  &0.0246&0.9012& &0.0247&0.8979& &0.0256&0.8956&\\
		&           &     interim &0.0052&0.3335& &0.0051&0.3252& &0.0051&0.3141&\\
		\hline
	\end{tabular}
	\vspace{5pt}
\end{table}
%\end{sidewaystable}
%%%%  end table 2  %%%%

%%%%  table 3  %%%%
%\begin{sidewaystable}
\begin{table}\centering
	\caption{Comparison of the correlations computed using different methods: the correlations calculated directly from the simulated samples ($\overline{cor}$), the predicted values using either stochastic process ($\widetilde{cor}_{sto}$) or exact distribution ($\widetilde{cor}_{exa}$), and the data-driven estimation ($\widehat{cor}$). The sample correlations were treated as the gold standard for a fair comparison on the other methods. In comparison with the gold-standard mean, correlations with difference $5-10\%$ were made \textit{italic}, and $>10\%$ were made \textbf{bold}. }
	\label{tab:corr}
	\begin{tabular}{lllcclccl}\hline
		\phantom{}&{}&\phantom{}&\multicolumn{2}{c}{ No violation} &
		&\multicolumn{2}{c}{ Violations I\&II} &\phantom{}\\ [-1pt]
		\cline{4-5} \cline{7-8}\\[-10pt]
		&Correlation pair& &$H_0$&$H_1$& &$H_0$&$H_1$& \\ \hline
		&$\mathcal{G}_{0,1}(t_{int})$&$\overline{cor} $&0.8329&0.8348&	&0.8261&0.8263&\\			
		& \ \&$\mathcal{G}_{0,0}(t_{int})$& $\widetilde{cor}_{sto}-\overline{cor} $ &-0.0029&-0.0038&	&0.0040&0.0047&\\
		&    & $\widetilde{cor}_{exa}-\overline{cor} $ &-0.0009&-0.0015&	&0.0059&0.0070&\\			
		&    & $\widehat{cor}-\overline{cor} $ &-0.0017&-0.0027&	&-0.0007&0.0006&\\
		
		&$\mathcal{G}_{0,1}(t_{fin})$&$\overline{cor} $&0.8452&0.8516&	&0.8482&0.8529&\\
		&  \ $\&\mathcal{G}_{0,0}(t_{fin})$  & $\widetilde{cor}_{sto}-\overline{cor} $ &0.0107&0.0145&	&0.0176&0.0230&\\
		&    & $\widetilde{cor}_{exa}-\overline{cor} $ &0.0120&0.0162&	&0.0188&0.0247&\\
		&    & $\widehat{cor}-\overline{cor} $ &-0.0010&-0.0018&	&-0.0008&-0.0014&\\
		
		&$\mathcal{G}_{0,0}(t_{int})$&$\overline{cor} $&0.7766&0.7791&	&0.7763&0.7797&\\
		&  \ $\&\mathcal{G}_{0,0}(t_{fin})$ & $\widetilde{cor}_{sto}-\overline{cor} $ &-0.0037&-0.0033&	&-0.0035&-0.0039&\\
		&    & $\widetilde{cor}_{exa}-\overline{cor} $ &-0.0046&-0.0016&	&-0.0043&-0.0021&\\
		&    & $\widehat{cor}-\overline{cor} $ &-0.0020&-0.0045&	&-0.0017&-0.0051&\\
		
		&$\mathcal{G}_{0,1}(t_{int})$&$\overline{cor} $&0.6457&0.6267&	&0.6158&0.5884&\\
		&    \ $\&\mathcal{G}_{0,1}(t_{fin})$& $\widetilde{cor}_{sto}-\overline{cor} $ &-0.0068&-0.0074&	&0.0232&\textit{0.0309}&\\
		&    & $\widetilde{cor}_{exa}-\overline{cor} $ &-0.0068&-0.0052&	&0.0232&\textit{0.0331}&\\ 
		&    & $\widehat{cor}-\overline{cor} $ &-0.0012&-0.0077&	&0.0000&-0.0082&\\
		
		&$\mathcal{G}_{0,1}(t_{int})$&$\overline{cor} $&0.6475&0.6493&	&0.6414&0.6454&\\ 
		&   \ $\&\mathcal{G}_{0,0}(t_{fin})$ & $\widetilde{cor}_{sto}-\overline{cor} $ &-0.0060&-0.0046&	&0.0002&-0.0008&\\
		&    & $\widetilde{cor}_{exa}-\overline{cor} $ &-0.0052&-0.0014&	&0.0009&0.0024&\\
		&    & $\widehat{cor}-\overline{cor} $ &-0.0037&-0.0047&	&-0.0020&-0.0049&\\             
		
		&$\mathcal{G}_{0,0}(t_{int})$&$\overline{cor} $&0.5375&0.5262&	&0.5093&0.4887&\\
		&    \ $\&\mathcal{G}_{0,1}(t_{fin})$& $\widetilde{cor}_{sto}-\overline{cor} $ &-0.0071&-0.0115&	&0.0211&\textit{0.0260}&\\
		&    & $\widetilde{cor}_{exa}-\overline{cor} $ &-0.0059&-0.0083&	&0.0223&\textit{0.0292}&\\
		&    & $\widehat{cor}-\overline{cor} $ &-0.0018&-0.0111&	&-0.0011&-0.0089&\\
		\hline
	\end{tabular}
	\vspace{5pt}
\end{table}
%\end{sidewaystable}
%%%%  end table 3  %%%%

\begin{figure}
	\centerline{
		\includegraphics[width=0.5\columnwidth]{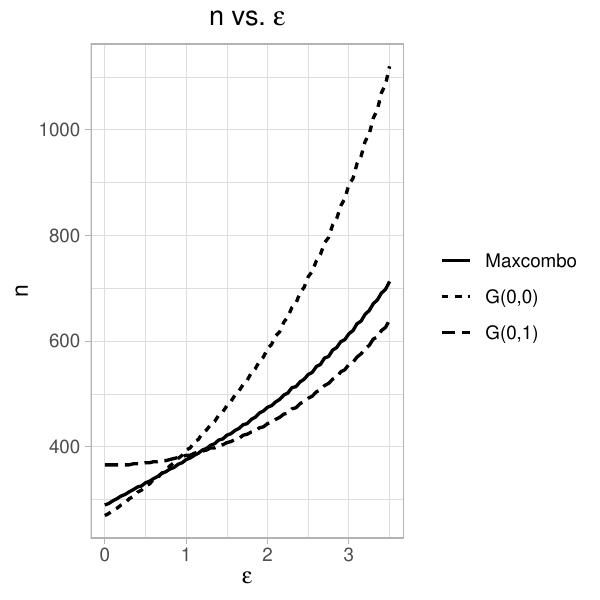}
		\includegraphics[width=0.5\columnwidth]{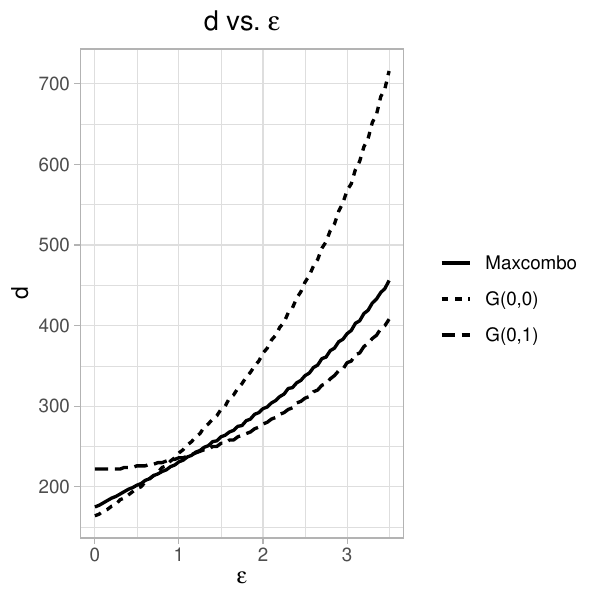}
	}
	\caption{Sample sizes needed to achieve the required power $\beta=0.9$ for the group sequential design in the presence of different delayed effect times ($\epsilon$). The required number of subjects ($n$) and events ($d$) were plotted against the delay-effect times $\epsilon$ in $[0,3.5]$. }
	\label{fig:2}
\end{figure}

To demonstrate the advantage of using maxcombo in group sequential designs, we computed the required sample sizes following (\ref{equ:beta2}) in comparison with all the single tests from the combo. The results were presented in Figure~\ref{fig:2} in terms of patient numbers ($n$) and event counts ($d$). The settings were consistent with those of simulations for Table~\ref{tab:default}: no assumption violation, fixing post-delay hazard ratio to be $\theta=0.6$ and varying the delayed time $\epsilon$ from $0$ to $3.5$. We plotted three curves in Figure~\ref{fig:2}: the GS-MC test of $\mathcal{G}_{max}(t)=\max(\mathcal{G}_{0,0}(t),\mathcal{G}_{0,1}(t))$, GS-WLRT ($\mathcal{G}_{0,1}(t)$) and GS-SLRT ($\mathcal{G}_{0,0}(t)$). According to Figure~\ref{fig:2}, when the delay-time is close to 0 ($<0.75$), $\mathcal{G}_{0,0}(t)$ requires the smallest sample size, while $\mathcal{G}_{0,1}(t)$ requires the biggest, consistent to the case that PH is dominant. When the waiting time before treatment effect $\epsilon$ is long ($>1.15$), $\mathcal{G}_{0,1}(t)$ becomes more powerful and thus requires a smaller sample size. Interestingly, in the intermediate state, when $\epsilon\in[0.75,1.15]$, the maxcombo test requires smaller sample sizes than the other two. It implies that the sample size needed from a maxcombo test is always nearly the most powerful one among all the tests in the combo. Therefore, we conclude that employing a maxcombo test in a group sequential design tends to reduce the sample size and largely improve the testing robustness. 

\section{Discussion}
\label{s:discuss}

In this report, we proposed a general framework for a group sequential design with maxcombo tests or GS-MC in short. The proposed design is completely simulation-free, can effectively control the type I error and find the required sample size to achieve the nominal ideal power under $H_1$. We have developed two prediction methods that are based on the assumed distributions under the two hypotheses ($H_0$ and $H_1$), and one data-driven estimation method. We demonstrated in our simulation studies that the proposed approaches displayed strong robustness towards various violations of assumptions related to survival, censoring, and accrual processes. 

Note that among the prediction methods, the stochastic approach can adapt to various survival functions. One can also extend the use of the exact prediction by repeating the derivations of the formulas in Appendix~B. Most often, the stochastic prediction method and the exact prediction method exhibit similar performance (depending on the number of intervals $b$), though the former is usually more flexible but also more computationally expensive than the latter. 

When we were preparing the manuscript, we notice another group also developed a correlation matrix approximation method, which shares some similar features to our proposal \cite{roychoudhury2019robust}. Their proposed method directly calculates all the correlations using the independent increment properties, but ours computes the third type of correlations based on the first two types. It can be shown that the two methods are numerically equivalent for the correlation matrix computation under $H_0$. Another major difference between the two methods lies in the fact that their approach is not entirely simulation-free, since their sample size calculation needs simulations.

We have established all the functions used in the proposed design in an R package on GitHub (lilywang1988/GSMC). Note that the proposed design can accommodate a regular maxcombo test without interim analysis by setting $M=1$. Although in this report, we restrict our scope to group sequential designs with sample sizes decided in advance, the proposed approach can be extended to adaptive designs, whereby sample sizes are adjusted based on the observed data.

%\backmatter
\iffalse
\section*{Acknowledgments}
This is acknowledgment text.\cite{Kenamond2013} Provide text here. This is acknowledgment text. Provide text here. This is acknowledgment text. Provide text here. This is acknowledgment text. Provide text here. This is acknowledgment text. Provide text here. This is acknowledgment text. Provide text here. This is acknowledgment text. Provide text here. This is acknowledgment text. Provide text here. This is acknowledgment text. Provide text here. 

\subsection*{Author contributions}

This is an author contribution text. This is an author contribution text. This is an author contribution text. This is an author contribution text. This is an author contribution text. 

\subsection*{Financial disclosure}

None reported.

\subsection*{Conflict of interest}

The authors declare no potential conflict of interests.

\fi

\section*{Supporting information}

The following supporting information is available as part of the online article:

\noindent
\textbf{Web Table~1}
{The rejection probabilities under the null hypothesis denoted by $H_0$ (type I error) and under the alternative hypothesis $H_1$ (power) when the censoring, accrual and event hazards are misspecified. }

\noindent
\textbf{Web Table~2}
{The rejection probabilities under the null hypothesis denoted by $H_0$ (type I error) and under the alternative hypothesis $H_1$ (power) when the censoring, accrual and delayed time are misspecified.}

\noindent
\textbf{Web Table~3}
{Comparison of the correlations computed using different methods \emph{without} violation in accrual (I), censoring (II), survival functions (III) or delayed time (IV).}

\noindent
\textbf{Web Table~4}
{Comparison of the correlations computed using different methods \emph{with} violations I (accrual) and II (censoring). }

\noindent
\textbf{Web Table~5}
{Comparison of the correlations computed using different methods \emph{with} violations I (accrual), II (censoring), and III(event rate). }

\noindent
\textbf{Web Table~6}
{Comparison of the correlations computed using different methods \emph{with} violations I (accrual), II (censoring), and IV(delayed time). }

\appendix

\section{Proof for Theorem~1}\label{app:sec:proof}
\begin{proof}
	This is equivalent to prove
	\begin{equation}\label{app:equ:proof_thm1}
	E[X_1X_2]E[X_2X_3]=E[X_1X_3]E[X_2^2]
	\end{equation}{}
	Since $X_3=X_2+M$ and $M\perp (X_1,X_2)$, we have the left-hand side (LHS) of (\ref{app:equ:proof_thm1}) to be 
	\begin{equation}
	\begin{array}{ll}
	LHS&=E[X_1X_2]E[X_2(\phi X_2+M)]\\
	&=\phi E(X_1X_2)E(X_2^2). 
	\end{array}
	\end{equation}{}
	The right-hand side (RHS) of (\ref{app:equ:proof_thm1}) is 
	\begin{equation}
	\begin{array}{ll}
	RHS&=E[X_1(\phi X_2+M)]E[X_2^2]\\
	&=\phi E(X_1X_2)E(X_2^2). 
	\end{array}
	\end{equation}{}
	Thus the equality in (\ref{app:equ:proof_thm1}) holds. 
\end{proof}{}

\section{Exact predictions} \label{app:sec:exact}
For simplicity, we demonstrated the exact prediction for a simple case following the survival functions given in Fine
\cite{fine2007consequences} and Hasegawa\cite{hasegawa2014sample}, where the control group is following an exponential distribution and the treatment group is following a a two-piece exponential with delayed effect $\theta$ at $\epsilon$: 

\begin{equation}\label{equ:survival_2p}
\begin{array}{cc}
S_0(s)= \exp(-\lambda s),   
& 
S_1(s)=\left\{\begin{array}{c l}
exp(-\lambda s) &for \ s\leq \epsilon,\\
c \exp(-\theta\lambda s)& for \ s> \epsilon;
\end{array}  \right.
\end{array}
\end{equation} 

\begin{equation}\label{equ:density_2p}
\begin{array}{cc}
f_0(s)= \lambda\exp(-\lambda s),   
& 
f_1(s)=\left\{\begin{array}{c l}
\lambda\exp(-\lambda s)=\lambda S_1(s) &for \ s\leq \epsilon,\\
\theta\lambda c\exp(-\theta\lambda s)=\theta\lambda S_1(s)& for \ s> \epsilon,
\end{array}  \right.
\end{array}
\end{equation} Note that in (\ref{equ:survival_2p}) and (\ref{equ:density_2p}), we have $c=\exp(-(1-\theta)\lambda \epsilon)$ and $\theta$ as the effective post-delay hazard ratio $\theta\in (0,1)$. If $\epsilon=0$, (\ref{equ:survival_2p}) and (\ref{equ:density_2p}) are reduced to a PH case.

According to \cite{hasegawa2016group}, the predicted variance or information is
\begin{equation}\label{equ:maxinformation_fomula}
\widetilde{V}_{exa}(G_{\rho,\gamma}(t))=np(1-p)\left[\int_0^t \min(\frac{t-s}{R},1)S(s)^{2\rho}(1-S(s))^{2\gamma}f(s)ds \right],
\end{equation} Under the null hypothesis, $S(t)=S_0(t)$ and $f(t)=f_0(t)$, but under the alternative hypothesis,  $S(t)=pS_1(t)+(1-p)S_0(t)$ and $f(t)=pS_1(t)+(1-p)S_0(t)$, where p is the treatment assignment probability. The major difference between $\widetilde{V}_{exa}$ and $\widetilde{V}_{sto}$ is that the at-risk proportion for the treatment group is fixed to be $p$ in $\widetilde{V}_{exa}(G_{\rho,\gamma}(t))$. 

We introduce some utility functions $uv()$, $u()$ and $v()$: 
\begin{equation}
\begin{array}{ll}
v(t,\epsilon,k,\lambda)&=I(\epsilon\leq t-R)\int_0^t\min(\frac{t-x}{R},1)\exp(-k\lambda x)\lambda I(x\leq\epsilon)dx\\
&=\frac{1}{k}\left\{1-\exp(-k\lambda\epsilon) \right\}
\end{array}
\end{equation}

\begin{equation}
\begin{array}{ll}
u(t,\epsilon,k,\lambda)&=I(\epsilon> t-R)\int_0^{t}\min(\frac{t-x}{R},1)\exp(-k\lambda x)\lambda I(x\leq\epsilon)dx\\
&=\frac{1}{k}\left[1-\exp(-k\lambda(t-R)^+)\right]+\frac{R\wedge t}{kR}\exp(-k\lambda(t-R)^+)\\
&\ -\frac{(t-\epsilon)^+}{kR}\exp(-k\lambda\epsilon\wedge t)+\frac{1}{k^2R\lambda}\left[\exp(-k\lambda\epsilon\wedge t)-\exp(-k\lambda(t-R)^+) \right],
\end{array}
\end{equation} where $(a)^+=\max(a,0)$ and $a\wedge b=\min(a,b)$.

\begin{equation}
uv(t,\epsilon,k, \lambda)= I(\epsilon>t-R) u(t,\epsilon,k,\lambda)+I(\epsilon\leq t-R) v(t,\epsilon,k,\lambda)
\end{equation}

Based on the basic utility functions, there are some other advanced utility functions, $h_1()$, $h_0()$ and $\widetilde{h}()$, for the variance/information prediction under the alternative hypothesis: 
\begin{equation}
\begin{array}{ll}
h_1(t,k_1,k_2)&=-\int_0^t \min(\frac{t-x}{R},1)S_1(x)^{k_1}S_0(x)^{k_2}dS_1(x)\\
&=uv(t,\epsilon,k_1+k_2+1,\lambda)+\\
&\hspace{1in}c^{k_1+1}\theta\left\{u(t,t,\theta(k_1+1)+k_2,\lambda)-uv(t,\epsilon,\theta(k_1+1)+k_2,\lambda)
\right\}
\end{array}
\end{equation}

\begin{equation}
\begin{array}{ll}
h_0(t,k_1,k_2)&=-\int_0^t \min(\frac{t-x}{R},1)S_1(x)^{k_1}S_0(x)^{k_2}dS_0(x)\\
&=uv(\tilde{t},k_1+k_2+1,\lambda)+\\
&\hspace{1in}c^{k_1}\left\{u(t,t,\theta k_1+k_2+1,\lambda)-uv(t,\epsilon,\theta k_1+k_2+1,\lambda)
\right\}
\end{array}
\end{equation}

\begin{equation}
\begin{array}{ll}
\tilde{h}(t,k)&=-\int_0^t\min(\frac{t-x}{R},1)S^k(x)dS(x)\\
&=-\int_0^t\min(\frac{t-x}{R},1)\sum_{i=0}^k {{k}\choose{i}}p^i(1-p)^{k-i}S_1^i(x)S_0^{k-i}(x)d[pS_1(x)+(1-p)S_0(x)]\\
&=\sum_{i=0}^k {{k}\choose{i}}p^{i+1}(1-p)^{k-i}h_1(t,i,k-i)+\sum_{i=0}^k {{k}\choose{i}}p^{i}(1-p)^{k-i+1}h_0(t,i,k-i)
\end{array}
\end{equation}

The the predicted variance under the null hypothesis using the exact assumed distributions is
\begin{equation}\label{maxinfo}
\begin{array}{l}
\widetilde{V}_{exa}(G_{\rho,\gamma}(t))=np(1-p)\times\left\{\begin{array}{c l}
u(t,t,1,\lambda)    &for \ \rho=0, \gamma=0;\\
u(t,t,3,\lambda)    & for \ \rho=1,\gamma=0;\\
u(t,t,1,\lambda)+u(t,t,3,\lambda)-2u(t,t,2,\lambda) &for \ \rho=0,\gamma=1;\\
u(t,t,3,\lambda)+u(t,t,5,\lambda)-2u(t,t,4,\lambda) &for \ \rho=1, \gamma=1. 
\end{array}  \right.
\end{array}
\end{equation}

The the predicted variance under the alternative hypothesis using the exact assumed distributions is
\begin{equation}
\begin{array}{l}
\widetilde{V}_{exa}(G_{\rho,\gamma}(t))=np(1-p)\times\left\{\begin{array}{c l}
\tilde{h}(t,0)    &for \ \rho=0, \gamma=0;\\
\tilde{h}(t,2)    & for \ \rho=1,\gamma=0;\\
\tilde{h}(t,0)+ \tilde{h}(t,2)-2 \tilde{h}(t,1) &for \ \rho=0,\gamma=1;\\
\tilde{h}(t,2)+\tilde{h}(t,4)-2 \tilde{h}(t,3) &for \ \rho=1, \gamma=1. 
\end{array}  \right.
\end{array}
\end{equation}

The predicted covariance values $\widetilde{Cov}_{exa}(\mathcal{G}_{\rho_1,\gamma_1}(t),\mathcal{G}_{\rho_2,\gamma_2}(t))$ can be obtained following the similar strategy using $u()$ and $\widetilde{h}()$ functions, by changing the squared weight of \ref{equ:maxinformation_fomula} to $S(t)^{\rho_1+\rho_2}(1-S(t))^{\gamma_1+\gamma_2}$. 

The formulas above can be easily extended to accommodate survival functions that are piece-wise exponential distributions with more than two pieces.

\bibliography{GSMC_sanofi.bib}%

\begin{thebibliography}{10}
\providecommand \doibase [0]{http://dx.doi.org/}%

\bibitem{reck2016pembrolizumab}
Reck M, Rodr{\'\i}guez-Abreu D, Robinson AG, et al. Pembrolizumab versus
  chemotherapy for PD-L1--positive non--small-cell lung cancer. {\it New
  England Journal of Medicine} 2016\string; 375(19)\string: 1823--1833.

\bibitem{mok2019pembrolizumab}
Mok TS, Wu YL, Kudaba I, et al. Pembrolizumab versus chemotherapy for
  previously untreated, PD-L1-expressing, locally advanced or metastatic
  non-small-cell lung cancer (KEYNOTE-042): a randomised, open-label,
  controlled, phase 3 trial. {\it The Lancet} 2019\string; 393(10183)\string:
  1819--1830.

\bibitem{mick2015statistical}
Mick R, Chen TT. Statistical challenges in the design of late-stage cancer
  immunotherapy studies. {\it Cancer immunology research} 2015\string;
  3(12)\string: 1292--1298.

\bibitem{harrington1982class}
Harrington DP, Fleming TR. A class of rank test procedures for censored
  survival data. {\it Biometrika} 1982\string; 69(3)\string: 553--566.

\bibitem{gehan1965generalized}
Gehan EA. A generalized Wilcoxon test for comparing arbitrarily singly-censored
  samples. {\it Biometrika} 1965\string; 52(1-2)\string: 203--224.

\bibitem{tarone1977distribution}
Tarone RE, Ware J. On distribution-free tests for equality of survival
  distributions. {\it Biometrika} 1977\string; 64(1)\string: 156--160.

\bibitem{schoenfeld1983sample}
Schoenfeld D. Sample-size formula for the proportional-hazards regression
  model. {\it Biometrics} 1983\string; 39(2)\string: 499--503.

\bibitem{schoenfeld1981asymptotic}
Schoenfeld D. The asymptotic properties of nonparametric tests for comparing
  survival distributions. {\it Biometrika} 1981\string; 68(1)\string: 316--319.

\bibitem{lee1996some}
Lee JW. Some versatile tests based on the simultaneous use of weighted log-rank
  statistics. {\it Biometrics} 1996\string: 721--725.

\bibitem{hasegawa2014sample}
Hasegawa T. Sample size determination for the weighted log-rank test with the
  Fleming--Harrington class of weights in cancer vaccine studies. {\it
  Pharmaceutical statistics} 2014\string; 13(2)\string: 128--135.

\bibitem{luo2019design}
Luo X, Mao X, Chen X, Qiu J, Bai S, Quan H. Design and monitoring of survival
  trials in complex scenarios. {\it Statistics in medicine} 2019\string;
  38(2)\string: 192--209.

\bibitem{lakatos1988sample}
Lakatos E. Sample sizes based on the log-rank statistic in complex clinical
  trials. {\it Biometrics} 1988\string: 229--241.

\bibitem{lee2007versatility}
Lee SH. On the versatility of the combination of the weighted log-rank
  statistics. {\it Computational Statistics \& Data Analysis} 2007\string;
  51(12)\string: 6557--6564.

\bibitem{fdaworkshop}
{Duke-Margolis Health Policy Center} . Public Workshop ``Oncology Clinical
  Trials in the Presence of Non-Proportional Hazards'': {
  $https://youtu.be/npufYAHeoxk$}.  {2018}.

\bibitem{hasegawa2016group}
Hasegawa T. Group sequential monitoring based on the weighted log-rank test
  statistic with the Fleming--Harrington class of weights in cancer vaccine
  studies. {\it Pharmaceutical statistics} 2016\string; 15(5)\string: 412--419.

\bibitem{tsiatis1981asymptotic}
Tsiatis AA. The asymptotic joint distribution of the efficient scores test for
  the proportional hazards model calculated over time. {\it Biometrika}
  1981\string; 68(1)\string: 311--315.

\bibitem{gordon1983discrete}
Gordon~Lan K, DeMets DL. Discrete sequential boundaries for clinical trials.
  {\it Biometrika} 1983\string; 70(3)\string: 659--663.

\bibitem{mvtnorm}
mvtnorm . https://cran.r-project.org/web/packages/mvtnorm/index.html (Last
  accessed Sept 20, 2019; version: 1.0.11).  2019.

\bibitem{roychoudhury2019robust}
Roychoudhury S, Anderson KM, Ye J, Mukhopadhyay P. Robust Design and Analysis
  of Clinical Trials With Non-proportional Hazards: A Straw Man Guidance from a
  Cross-pharma Working Group. {\it arXiv preprint arXiv:1908.07112} 2019.

\bibitem{fine2007consequences}
Fine GD. Consequences of delayed treatment effects on analysis of time-to-event
  endpoints. {\it Drug information journal} 2007\string; 41(4)\string:
  535--539.

\end{thebibliography}

\clearpage
\iffalse
\section*{Author Biography}

\begin{biography}{\includegraphics[width=66pt,height=86pt,draft]{empty}}{\textbf{Author Name.} This is sample author biography text this is sample author biography text this is sample author biography text this is sample author biography text this is sample author biography text this is sample author biography text this is sample author biography text this is sample author biography text this is sample author biography text this is sample author biography text this is sample author biography text this is sample author biography text this is sample author biography text this is sample author biography text this is sample author biography text this is sample author biography text this is sample author biography text this is sample author biography text this is sample author biography text this is sample author biography text this is sample author biography text.}
\end{biography}
\fi
\end{document}